\documentstyle[12pt,a4]{article}
\begin{document}
\thispagestyle{empty}
\begin{center}
\LARGE Diluted neural networks with adapting and correlated
synapses

~\\
~\\

\vspace{1.cm} \normalsize Massimo Mannarelli, Giuseppe Nardulli,
and Sebastiano Stramaglia\\ \vspace{0.5cm} {\it Center of
Innovative Technologies for Signal Detection and Processing, Bari}
\\ {\it Dipartimento di Fisica, I.N.F.N. Sezione di Bari\\ via
Amendola 173, 70126 Bari, Italy}\\ ~\\ ~\\
\end{center}
\vspace{1.cm}
\begin{abstract}
We consider the dynamics of diluted neural networks with clipped
and adapting synapses. Unlike previous studies, the learning rate
is kept constant as the connectivity tends to infinity: the
synapses evolve on a time scale intermediate between the quenched
and annealing limits and all orders of synaptic correlations must
be taken into account. The dynamics is solved by mean-field
theory, the order parameter for synapses being a function. We
describe the effects, in the double dynamics, due to synaptic
correlations.
\end{abstract}

\vspace{1.cm} \noindent PACS numbers: 87.10.+e, 05.20.-y

\vskip .5cm

In the past years, many models with a coupled dynamics of fast
Ising spins and slow interactions have been studied to understand
the simultaneous learning and retrieval in recurrent neural
networks \cite{coolen,shino}. A major approach to this problem is
replica mean-field theory with the replica number being the ratio
of two temperatures characterizing the stocasticity in the spin
dynamics and the interaction dynamics, respectively
\cite{cps,dot}. Recently this approach has been used to study
coupled dynamics in the XY spin glass \cite{bolle,bolle1}; the
generalization of these ideas \cite{vanm} to the case of a
hierarchy of subsystems with different characteristic time-scales,
in the Sherrington-Kirckpatrick model, interestingly leads to
Parisi's solution \cite{parisi}. Other approaches to coupled
dynamics in neural networks are described in \cite{heerema}, using
a Discrete Time Master Equation approach, and in \cite{roberts},
exploring {temporal learning rules}. Stochastic learning rules in
diluted neural networks were considered in \cite{lattanzi}: it was
shown that in order to preserve the associative memory capability
of the network the learning rate $q$ must be kept very small
(e.g., $q=O(1/K)$, where $K$ is the connectivity). Moreover, in
\cite{lattanzi}  the choice of a very small learning rate implied
that the correlation between synaptic variables could be neglected
so that the dynamics was solved by flow equations for a few number
of macroscopic order parameters. It is the purpose of this work to
reconsider coupled dynamics in diluted neural networks and keep
the learning rate fixed as the connectivity $K$ tends to $\infty$.
The dynamics of the network, in this limit, can be exactly solved
by taking into account all the orders of correlations between
synapses, the order parameter for synapses being a function on the
interval $[-1,1]$. According to the argument in \cite{lattanzi},
the functioning of this model as an associative memory is
questionable; we regard it as a simple model to analyze the
effects due to synaptic correlations in the double dynamics.

As in \cite{lattanzi} we consider a diluted neural network with
uni-directional synapses obeying a stochastic learning mechanism
\cite{amit}. The model is made of $N$ three states neurons
$s_i=0,\pm 1$, each connected (by binary synapses $J_{ij}=\pm 1$)
to $K$ input sites, chosen at random among the $N$ sites. The
parallel rule for updating synapses is the following: with
probability $q$ each synapse  $J_{ij}$ assumes the value $s_i s_j$
if this product is not zero; otherwise the synapse remains
unchanged. A parallel stochastic dynamics with inverse temperature
$\beta$ is assumed for neurons, where the local field acting on
neuron $s_i$ is given by $h_i=(\sum J_{ij} s_j)/K$, the sum being
over the input neurons. The coupled dynamics consists in alternate
updating of neurons and synapses. We will consider the limit
$N,K\to\infty$ with $K << ln \;N$: it is well known \cite{zip}
that neurons can then be treated as i.i.d. stochastic variables.
Moreover we choose $q$ constant as $K\to\infty$: $q$ controls the
ratio between the time scales over which neurons and synapses
evolve and the {\it adiabatic} approximation is recovered by
sending $q$ to zero \cite{caticha}. As a consequence, in the
present case one can not neglect the correlations among synapses.

Let us denote $s_1$, $s_2$, .....,$s_K$ the input neurons and
$J_1$, $J_2$, ...., $J_K$ the set of $K$ input synapses for a
given neuron $s_0$ (due to the translational symmetry the
following reasoning holds for an arbitrary $s_0$).

We start considering the following simple situation: the synapses
being independently updated by the transition matrix:
$
T\left({\bf J}|{\bf J'}\right)= \prod_{\alpha=1}^{K}
\tau\left(J_\alpha|J'_\alpha\right) $, where the transition matrix
for the single synapse is the following:

\[ \tau = \left( \begin{array}{cc}
              1-A & B \\
              A & 1-B
    \end{array}\right) \]
A good order parameter for synapses is $x=(\sum^K_{\alpha =1}
J_\alpha)/K\; \in [-1,1]$. Indeed, denoting with $\rho_t (x)$ the
pdf for $x$ at time $t$, one can demonstrate (see the Appendix)
that in the large $K$ limit the evolution of $x$ is ruled by a
deterministic Liouville operator:
\begin{equation}
\rho_{t+1} (x)= \int_{-1}^1 dy\delta \left(x-\hat{x}(y)\right)
\rho_t (y)
\label{eq:19}
\end{equation}
with $\hat{x}=B-A+y(1-A-B)$. The moments of $\rho_t$ provide the
synaptic correlations:
\begin{equation}
\langle x^p\rangle_t =\int dx\; x^p \rho_t (x)= \langle
J^{(1)}J^{(2)}\ldots J^{(p)}\rangle_t,
\end{equation}
where the synapses $J^{(1)}$, $J^{(2)}$, $\ldots J^{(p)}$ are all
different. The probability distribution, at time $t$, for the
local field acting on neuron $s_0$ is
\begin{equation}
P_t(h)={1\over m_t}\; \rho_t \left({h\over m_t}\right)\;,\; h\in
[-m_t,m_t], \label{eq:21}
\end{equation}
where $m_t$ is $\langle s\rangle_t$, the average magnetization of
the neuronic configuration. We will denote $Q_t=\langle s^2
\rangle_t$ the activity of neurons, satisfying $Q_t\ge m_t$ for
every time $t$.

Let us now come back to our problem. Due to the synaptic learning
rules, the values of $A$ and $B$ now depend on the value of $s_0$.
If $s_0 =0$ then $A=B=0$ and $\hat{x}=y$. If $s_0=1$ then
$A=q\left({Q_t-m_t\over 2}\right)$, $B=q\left({Q_t+m_t\over
2}\right)$ and $\hat{x}=qm_t+y\left(1-qQ_t\right)$. If $s_0=-1$
then $A=q\left({Q_t+m_t\over 2}\right)$, $B=q\left({Q_t-m_t\over
2}\right)$ and $\hat{x}=-qm_t+y\left(1-qQ_t\right)$. This implies
that even if at time $t$ we know $x$ exactly (i.e., $\rho_t$ is a
$\delta$-function), at time $t+1$  $x$ is not determined
($\rho_{t+1}$ will generically be a convex sum of three
$\delta$'s). The full distribution $\rho$ now plays the role of
order parameter for the synaptic variables, the time evolution law
being given by a mixture of three Liouville operators:
\begin{equation}
\begin{array}{cr}
\rho_{t+1}(x)&=(1-Q_t)\rho_t(x)+ {Q_t+m_t\over 2\left(1-qQ_t
\right)}\theta\left(1-\left|{x-qm_t\over 1-qQ_t}
\right|\right)\rho_t\left({x-qm_t\over
1-qQ_t}\right)\\&+{Q_t-m_t\over
2\left(1-qQ_t\right)}\theta\left(1- \left|{x+qm_t \over
1-qQ_t}\right|\right)\rho_t\left({x+qm_t\over
 1-qQ_t}\right);
\end{array}
\label{evol}
\end{equation}
$\theta $ is Heaviside's function.

Let us now consider the dynamics of neurons. We assume the
following form for the conditional probability for neurons:
\begin{equation}
P({\bf s_{t+1}}|h)\propto \exp{\beta(h{\bf s_{t+1}}+a{\bf
s^2_{t+1}})}, \label{rr}
\end{equation}
where ${\bf s_t}$ is the vector of neurons at time $t$, and $a$
controls the mean activity of the network. The time evolution law
for neuronic order parameters is then given by
\begin{equation}
\begin{array}{cc}
m(t+1)=&\int_{-1}^1 \;dx \rho_t (x) {2 sinh(\beta x m_t)\over 2
cosh(\beta x m_t) +e^{-\beta a}}\\ Q(t+1)=&\int_{-1}^1 \;dx \rho_t
(x) {2 cosh(\beta x m_t)\over 2 cosh(\beta x m_t) +e^{-\beta a}}.
\end{array}
\label{neu}
\end{equation}
These two equations, together with (4) and the initial conditions,
$m_0$, $Q_0$ and $\rho_0 (x)$, solve the double dynamics for the
present model.

Now we turn to analyze the flow equations. Firstly we consider the
case of $m$ and $Q$ being kept constant: $\rho_t$ tends
asymptotically to the invariant distribution $\rho_\infty$ of (4).
One can easily derive a recurrence formula for the moments of the
stationary distribution:

\begin{equation}
\langle x^n \rangle_\infty= \sum^{'}
\left(\begin{array}{c}n\\k\end{array}\right) \left(1-qQ
\right)^{n-k}\left(qm\right)^k \langle x^{n-k}\rangle_\infty+
{m\over Q}\sum^{''} \left(\begin{array}{c}n\\k\end{array}\right)
\left(1-qQ\right)^{n-k}\left(qm\right)^k \langle
x^{n-k}\rangle_\infty
\end{equation}

where $\sum^{'}$ ($\sum^{''}$) is over even (odd) positive
integers less than or equal to $n$. The invariant distribution is
a $\delta$ -function in the following cases. If $m=0$ then
$\rho_\infty =\delta (x)$. If $m=\pm 1$ then
$\rho_\infty=\delta(x-1)$, and in the adiabatic limit $q\to 0$ we
have $\rho_\infty \to\delta (x-m^2/Q^2)$. In the general case the
first two cumulants are given by:
\begin{equation}
\langle x\rangle_\infty ={m^2\over Q^2}
\end{equation}
which is independent of $q$, and
\begin{equation}
\langle x^2\rangle_\infty - \langle x\rangle_\infty^2={q\over
2-qQ}\left({m^2\over Q}-{m^4\over Q^3}\right).
\end{equation}

The last formula clearly shows how the synaptic correlations are
controlled by the learning rate $q$. For example, in Figure 1 the
invariant distribution of (4), we numerically find, is depicted
(for $q=0.06$, $Q=0.8$, and $m=0.5$). We compare it with the
$x$-distribution, over time, we find simulating a system of $K$
synapses, evolving by the stochastic learning mechanism, where
neurons $s_0$ and $\{s_\alpha \}$ are independently sampled with
$\langle s\rangle =m$ and $\langle s^2 \rangle =Q$ at each time
step. The agreement with the theoretical curve increases as $K$
grows and it is fairly good already for $K=500$ (see Fig. 1).

The stationary regime of the coupled dynamics shows a paramagnetic
phase with $m=0$ and a ferromagnetic phase with $m\ne 0$
\cite{caroppo}. By numerical analysis we find the transition line
between the two phases in the $\beta -a$ plane: in Figure 2 our
results are shown for some values of $q$. At fixed $a$, the
critical temperature decreases as $q$ is increased: the synaptic
correlations seem to amplify the disordering capability of thermal
noise. The two phases are separated by a first order transition,
in agreement with \cite{cps} where the para-ferro transition
changes from second to first order as the influence of spins on
the couplings dynamics becomes dominant.

Let us now study the role of adapting synapses in the damage
spreading phenomenon (see, e.g., \cite{derrida}). For simplicity
we assume two state neurons $s=\pm 1$, and we work in the
disordered phase $m=0$. We assume the local fields to be:
\begin{equation}
h_i = {1\over \sqrt{K}} \sum J_{ij} s_j +B_i
\end{equation}

where $B_i$ are random magnetic fields whose Gaussian distribution
has variance $B$, and the normalization has been chosen
differently from the previous case so as to have
 a non trivial $K\to\infty$ limit in this case.
We assume to be at zero temperature and consider two replicas of
the system, subject to the same random fields and the same noise
in the stochastic learning mechanism. We introduce the order
parameters $\Delta$ and $\epsilon$ defined as follows: ${1\over
2}(1+\Delta)$ is the probability that two corresponding synapses,
in the two replicas, are equal, while ${1\over 2}(1+\epsilon)$ is
the probability that two corresponding neurons, in the two
replicas, are equal. As in the previous section, one easily finds
that even if $\Delta$ is exactly known at a certain time, it is
not determined al later times: it must be described by a
probability distribution $\Gamma_t (\Delta )$, whose evolution is
given by eq.(4) with $Q=1$ and $m_t$ replaced by $\epsilon_t$.
While keeping fixed $\Delta$, the variables $\{Js\}$ are equal, in
the two replicas, with probability ${1\over 2}
(1+\Delta\epsilon)$. Therefore the local fields in the two
replicas can be written $h_1=X+Y$ and $h_2=X-Y$, where $X$ and $Y$
are random Gaussian variables with variance, respectively
$\sigma_X=(1+\Delta\epsilon)/2+B$ and
$\sigma_Y=(1-\Delta\epsilon)/2$. One can then easily obtain the
time evolution law for $\epsilon$:
\begin{equation}
\epsilon_{t+1}=1-{4\over \pi}\int_{-1}^1 d\Delta \Gamma_t (\Delta
)
 tan^{-1}\sqrt{{1-\Delta\epsilon_t\over 1+\Delta\epsilon_t+2B}}.
\end{equation}
Studying damage spreading is equivalent to check the stability of
the trivial fixed point $\epsilon =1$ and $\Gamma =\delta (\Delta
-1)$, corresponding to two identical replicas. We find that, for
every finite $B$, damage spreading occurs and a nontrivial fixed
point $\epsilon^{*} <1$ is stable. For low values of $q$ the
stationary distribution $\Gamma$ is peaked around its average
$\epsilon^2$: approximating the $tan^{-1}$ by Taylor expansion at
the second order around $\Delta =\epsilon^2$, the equation for the
fixed point reads:
\begin{equation}
\epsilon^{*}=1-{4\over \pi}tan^{-1}\sqrt{{1- \epsilon^{*3}\over
1+\epsilon^{*3}+2B}}+{{\cal C}\;B\epsilon^{*2}\over
\pi\left((1-\epsilon^{*3}) (1+\epsilon^{*3}+2B)\right)^{3\over
2}}\;,
\end{equation}
where ${\cal C}=\langle \Delta^2\rangle-\langle
\Delta\rangle^2=q(\epsilon^{*2} -\epsilon^{*4})/(2-q)$ at
equilibrium.

The solution $\epsilon^{*}$ of the equation above is the
asymptotic correlation between neurons in the two replicas as a
function of $q$. In Figure 3 we depict $\partial
\epsilon^{*}/\partial q_{|q=0}$ versus $B$. Since we find this
quantity to be always positive, it follows that the synaptic
correlations act against the damage spreading phenomenon and tend
to increase the correlation between the configurations of neurons
in the two replicas, as one might intuitively expect.

We have described an exactly solvable model of double dynamics
where synaptic correlations, arising from a stochastic learning
mechanism, are important at all orders. The order parameter for
synapses in the mean-field dynamical theory is a function whose
evolution is given by a mixture of Liouville operators. The
critical temperature for the ferromagnetic transition is found to
decrease as the learning rate increases: there is a wide range of
temperatures such that the system may order or not depending on
the speed at which it adapts, and ordering is asymptotically
achieved only if the adaptation is sufficiently slow. We also
outlined the role played by synaptic correlations in the damage
spreading phenomenon.

\section*{Appendix}
We show the validity of equation (\ref{eq:19}). Using the same
notation as in the text, let $P_t\left({\bf J}\right)$ be the pdf
for synapses at time $t$. Then
\begin{equation}
P_{t+1}\left({\bf J}\right)= Tr_{\bf J'} T\left({\bf J}|{\bf
J'}\right)
 P_t\left({\bf J'}\right).
\label{eq:2}
\end{equation}
It is useful to observe that, due to the symmetry of our problem,
the distribution $P_t\left({\bf J}\right)$ will be symmetric under
permutations of synapses (provided initial conditions respect the
symmetry). It follows that $P_t$ is a function of the only
non-trivial invariant for permutations one can build out of $K$
binary variables, i.e. $x={1\over K}\sum^K_{\alpha =1} J_\alpha$.

After standard calculations \cite{coolen}, the probability
distribution for $x$, $\rho_t (x)$, is found to evolve according
to
\begin{equation}
\rho_{t+1} (x)= \int_{-1}^1 dy W_K (x,y) \rho_t (y), \label{eq:6}
\end{equation}
where the time-independent kernel $W_K$ is given by
\begin{equation}
W_K(x,y)= {Tr_{\bf J} Tr_{\bf J'} \delta \left( y-{1\over K}\sum
J'\right)
  \delta \left( x-{1\over K}\sum J\right) \prod_{\alpha=1}^{K} \tau\left(J_\alpha|J'_\alpha
\right) \over Tr_{\bf J'} \delta \left( y-{1\over K}\sum J'\right)
}. \label{eq:8}
\end{equation}
The structure of this kernel is, in the limit $K\to\infty$:
\begin{equation}
W_K(x,y)= {K^2\over (2\pi i)^2}\int_{-i\infty}^{i\infty}d\lambda
\int_{-i\infty}^{i\infty}d\mu \;e^{K F(\lambda, \mu, x,y)},
\label{eq:16}
\end{equation}
where
\begin{equation}
F(\lambda, \mu, x,y)=L(\lambda, \mu)-S(y)-\lambda y -\mu x,
\label{eq:17}
\end{equation}
\begin{equation}
S(y)=-{1+y\over 2}log{1+y\over 2}-{1-y\over 2}log{1-y\over 2},
\label{eq:10}
\end{equation}
\begin{equation}
e^{L(\lambda,\mu)}=(1-A)e^{\lambda +\mu}+A e^{\lambda -\mu}+B
e^{-\lambda +\mu}+(1-B) e^{-\lambda
 -\mu};
\label{eq:15}
\end{equation}
the time evolution for the synaptic distribution is then given by
the following equation:
\begin{equation}
\rho_{t+1} (x)= {K^2\over (2\pi i)^2}\int_{-1}^1
dy\int_{-i\infty}^{i\infty}d\lambda \int_{-i\infty}^{i\infty}d\mu
\;e^{K F(\lambda,\mu,x,y)} \rho_t (y). \label{eq:18}
\end{equation}
As a consequence, in the large $K$ limit the integral in
(\ref{eq:18}) is dominated by the physical saddle point, this
means that the evolution operator $W$ becomes, in the large $K$
limit, a Liouville operator, describing a deterministic evolution.
The saddle point is determined by the equations: $ {\partial F/
\partial \lambda}=0,\;\;{\partial F/
\partial \mu} =0,\;\;{\partial F/ \partial y}=0$.
After a little algebra, it turns out that at the saddle point the
relation $x=B-A+y(1-A-B)$ holds. Since $W_K$ is (by construction)
normalized for every $K$, also the limiting kernel, as $K$ goes to
infinity, will be normalized: we can then conclude that the
limiting kernel is given by $\delta \left(x-\hat{x}(y)\right)$,
where $\hat{x}=B-A+y(1-A-B)$.

\newpage
\noindent\Large\textbf{Figure Captions} \normalsize \vspace{1.0cm}
\begin{description}
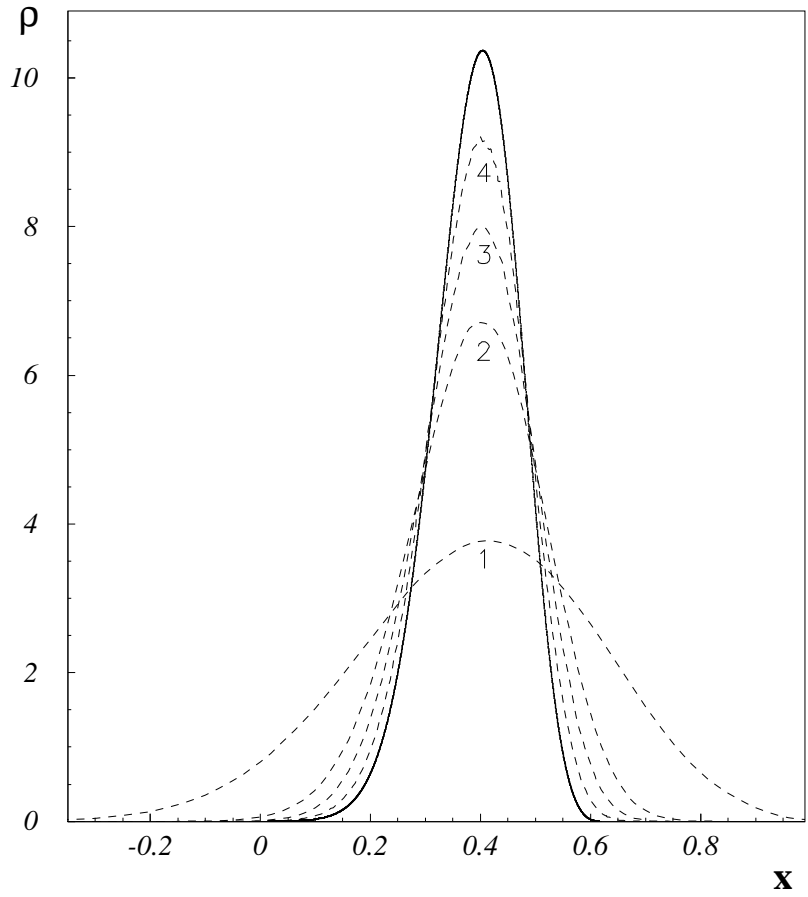
\item{Figure 1}: The dashed lines represent
the $x$-distributions from numerical simulations for $K=20$ (1),
$K=100$ (2), $K=200$ (3), $K=500$ (4), to be compared with the
invariant distribution of (4), here represented by the solid line.
The case $q=0.06$, $Q=0.8$ and $m=0.5$ is here considered.
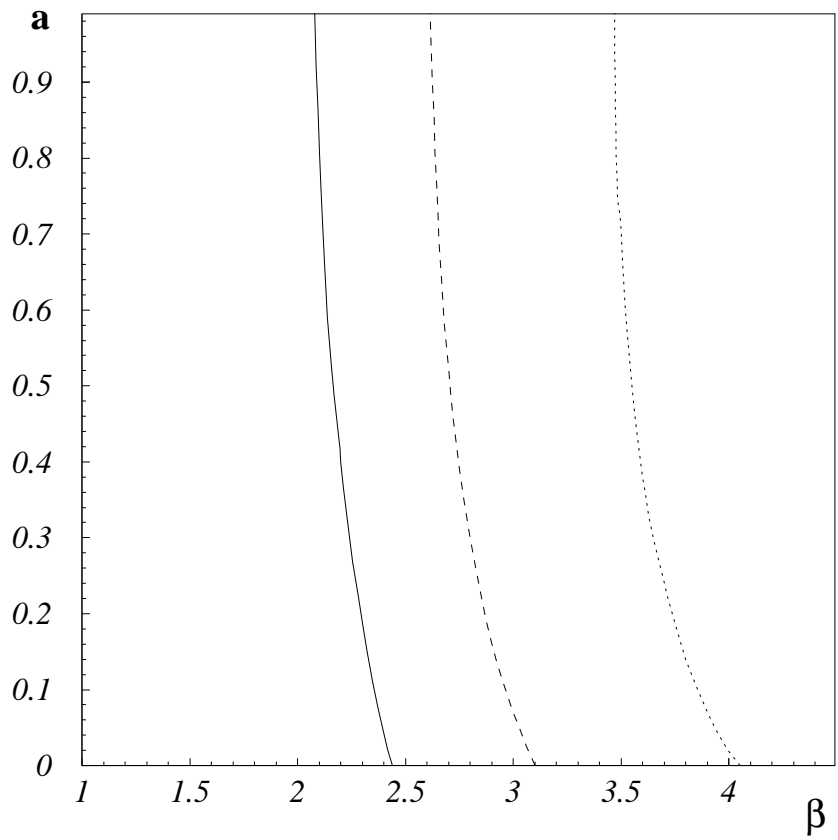
\item{Figure 2}:
In the plane $\beta -a$ of parameters (see the text), the
transition lines between the ferro and paramagnetic phases are
depicted, for $q=0$ (continuous line), $q=0.02$ (dashed line) and
$q=0.05$ (dotted line).
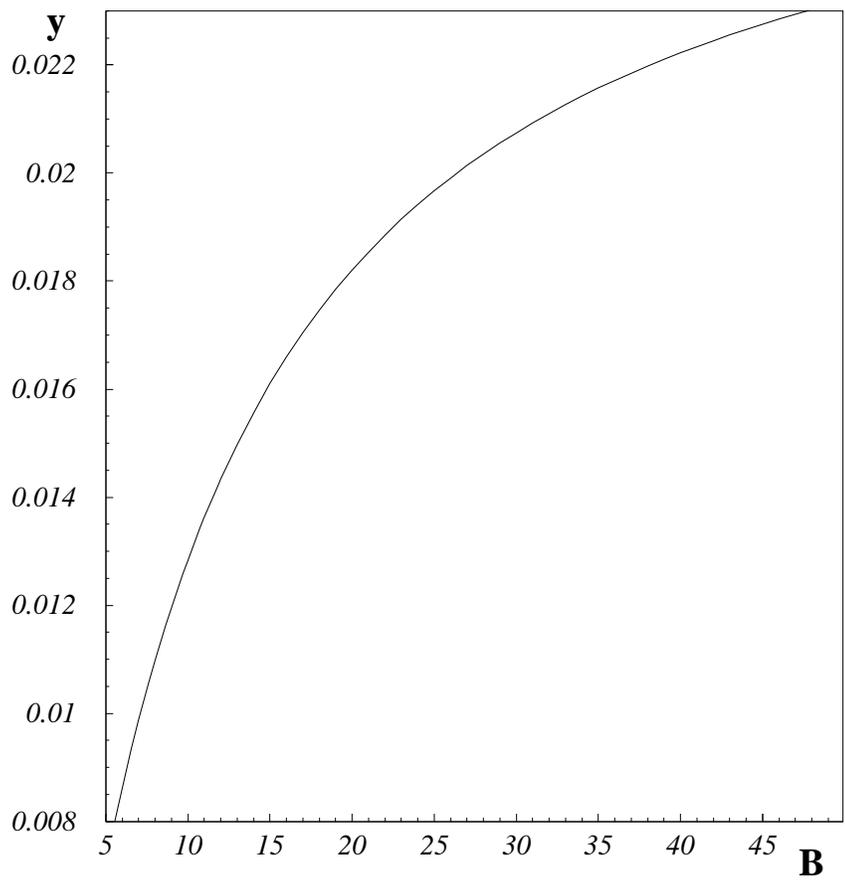
\item{Figure 3}:
Concerning the damage spreading phenomenon, $y=\partial
\epsilon^{*}/\partial q_{|q=0}$ is depicted versus the variance of
random fields, $B$ (see the text).

\end{description}
\end{document}